\documentclass[aps,pre,twocolumn,superscriptaddress,english,showpacs]{revtex4-1}
\usepackage[T1]{fontenc}
\usepackage[latin9]{inputenc}
\usepackage{amsmath}
\usepackage{graphicx}

\makeatletter

\providecommand{\tabularnewline}{\\}

\@ifundefined{showcaptionsetup}{}{
\PassOptionsToPackage{caption=false}{subfig}}
\usepackage{subfig}
\makeatother

\usepackage{babel}

\begin{document}

\title{How nonuniform contact profiles of T cell receptors modulate thymic
selection outcomes}

\author{Hanrong Chen}
\affiliation{Harvard John A. Paulson School of Engineering and Applied Sciences, Harvard University,
Cambridge, Massachusetts 02138, USA}
\author{Arup K. Chakraborty}
\affiliation{Departments of Chemical Engineering, Chemistry, and Biological Engineering, Institute for Medical Engineering and Science,
Massachusetts Institute of Technology, Cambridge, Massachusetts 02139, USA}
\affiliation{Department of Physics, Massachusetts Institute of Technology, Cambridge, Massachusetts 02139, USA}
\author{Mehran Kardar}
\affiliation{Department of Physics, Massachusetts Institute of Technology, Cambridge, Massachusetts 02139, USA}

\date{\today}

\begin{abstract}
T cell receptors (TCRs) bind foreign or self-peptides attached to
major histocompatibility complex (MHC) molecules, and the strength of
this interaction determines T cell activation. Optimizing the ability of T cells
to recognize a diversity of foreign peptides yet be tolerant of self-peptides
is crucial for the adaptive immune system to properly function.
This is achieved by selection of T cells in the thymus,
where immature T cells expressing unique, stochastically generated TCRs
interact with a large number of self-peptide-MHC; if a TCR does not bind
strongly enough to any self-peptide-MHC, or too strongly with at least one self-peptide-MHC,
the T cell dies. Past theoretical work cast thymic selection as an extreme value problem,
and characterized the statistical enrichment or depletion of amino acids in the post-selection
TCR repertoire, showing how T cells are selected to be
able to specifically recognize peptides derived from diverse pathogens, yet have limited self-reactivity.
Here, we investigate how the degree of enrichment is modified by nonuniform contacts that a TCR
makes with peptide-MHC.
Specifically, we were motivated by recent experiments showing that
amino acids at certain positions of a TCR sequence have large effects on thymic
selection outcomes, and crystal structure data that reveal a nonuniform
contact profile between a TCR and its peptide-MHC ligand. Using a representative TCR contact
profile as an illustration, we show via simulations that the degree of enrichment now varies by
position according to the contact profile, and, importantly, it depends on the
implementation of nonuniform contacts during thymic selection.
We explain these nontrivial results analytically. Our study has implications for
understanding the selection forces that shape the functionality of the post-selection
TCR repertoire.
\end{abstract}

\maketitle

\section{Introduction}

T cell receptors (TCRs) bind peptides loaded onto
major histocompatibility complex (MHC) molecules (abbreviated as peptide-MHC)
on the surface of antigen presenting cells (APCs), and the strength of
this interaction determines T cell activation~\cite{1996-Alam,2005-Krogsgaard-Davis}.
Such peptides are derived from either the host itself (self-antigens), or, potentially,
pathogens infecting the host (foreign antigens); thus, optimizing the ability of T cells to
recognize diverse foreign antigens with high specificity, yet be self-tolerant, is
essential for the proper functioning of the adaptive immune system.
Immature T cells (or thymocytes) express a distinct TCR on their surface
assembled through a stochastic process of gene rearrangement, generating
a highly diverse repertoire~\cite{2012-Mora-Walczak-Callan} that
is acted upon by selection in the thymus. There, thymocytes are screened
against a large number of self-peptide-MHC; those that do not productively
bind to self-peptide-MHC die (this is called positive selection),
and those that bind too strongly are also eliminated (this is called
negative selection). T cells that survive thymic selection are exported
to the body's periphery where they participate in the adaptive immune
response. In this way, thymic selection shapes the post-selection
TCR repertoire to potentially recognize diverse foreign antigens, yet
limit self-reactivity~\cite{2005-Huseby}. While detailed statistics
of the post-selection repertoire are now available
~\cite{2014-Callan-Mora-Walczak,2017-Callan-Mora-Walczak,2016-Stadinski},
how thymic selection achieves this outcome is less well-understood.

\begin{figure}
\begin{tabular}{cc}
\includegraphics[scale=0.27]{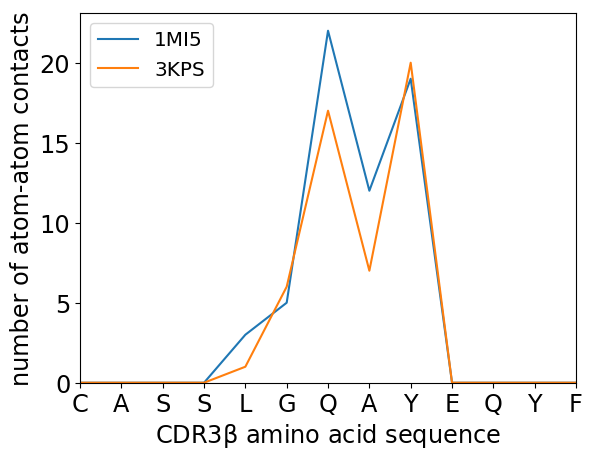} & \includegraphics[scale=0.27]{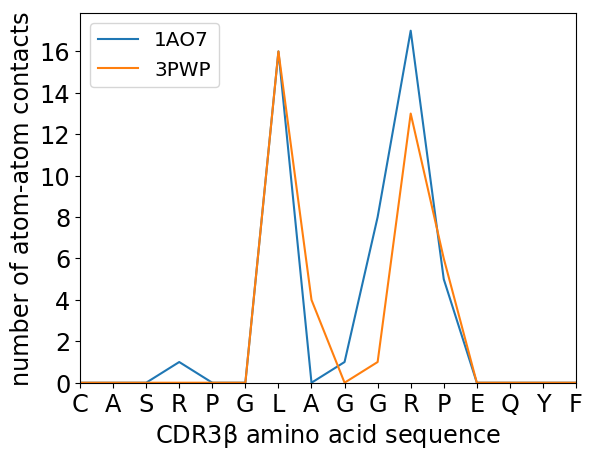}\tabularnewline
\includegraphics[scale=0.27]{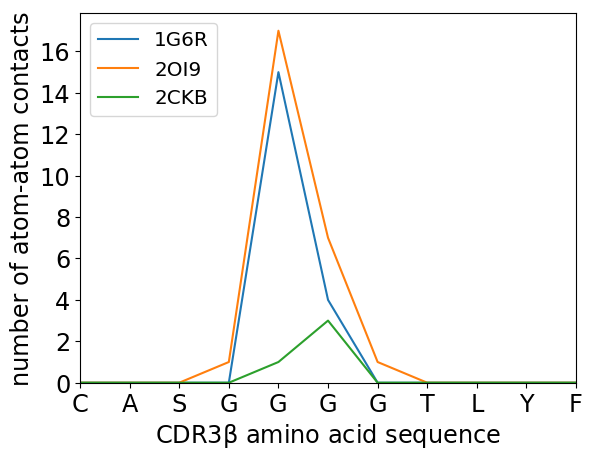} & \includegraphics[scale=0.27]{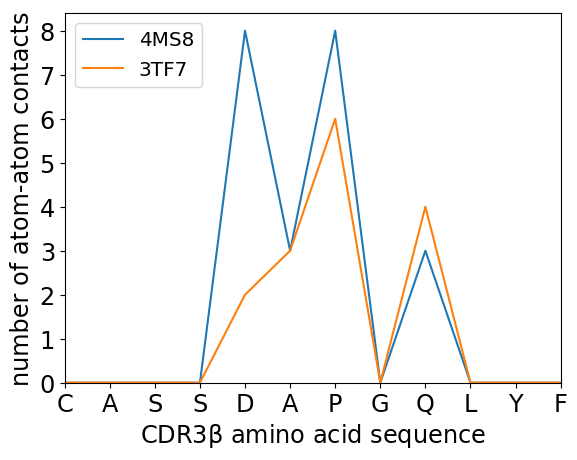}\tabularnewline
\end{tabular}

\caption{Number of atom-atom contacts (within 4Å) between TCR amino acids and
peptide-major histocompatibility complex (MHC) molecules, measured
from crystal structures of TCR\textendash peptide-MHC complexes (contact
data from~\cite{2016-Stadinski}; Protein Data Bank IDs are shown
on top left of figures). Each x-axis shows a single TCR$\beta$-chain
complementarity-determining region (CDR3$\beta$) amino acid sequence,
and each colored line plots the number of contacts made by that CDR3$\beta$
amino acid when bound to separate peptide-MHC molecules. These plots
are a representation of the nonuniform contact interface between a
TCR and its peptide-MHC ligand.\label{fig:Number-of-inter-atomic}}
\end{figure}

Mathematical models have shed light on various aspects of thymic selection
(for a review, see Ref.~\cite{2014-Yates}); in particular, models
that represented TCR\textendash peptide-MHC interactions as pairwise
interactions between digit strings~\cite{1999-Detours-Mehr-Perelson,1999-Detours-Perelson}
were useful for studying how TCR cross-reactivity can result. In
Refs.~\cite{2008-Kosmrlj,2009-Kosmrlj}, some of the authors considered
a more explicit representation of inter-amino acid interaction strengths~\cite{Miyazawa-Jernigan},
and characterized the statistical enrichment or depletion of amino acids in the
post-selection TCR repertoire, computationally~\cite{2008-Kosmrlj} and
analytically~\cite{2009-Kosmrlj}, as a function of parameters such as the
number of self-peptide-MHC encountered during selection.
These studies provided insight into how T cells are selected to be specific for unknown
foreign peptides and yet are self-tolerant.

In the past few years, more detailed information about thymic selection
outcomes have emerged. Advances in high-throughput sequencing have
allowed researchers to quantify the statistics of post-selection TCR
sequences in detail, revealing positional differences in the enrichment
of amino acids~\cite{2014-Callan-Mora-Walczak,2017-Callan-Mora-Walczak}.
Furthermore, recent experiments found that pre-selection thymocytes that were activated
by self-peptide-MHC (and hence would fail negative selection) were
enriched in hydrophobic amino acids at positions 6 and 7 of the TCR$\beta$-chain
complementarity-determining region (CDR3$\beta$), while
thymocytes passing both positive and negative selection were enriched
in amino acids with moderate hydrophobicity at these positions~\cite{2016-Stadinski}.
These results agree with theoretical predictions made in previous
work~\cite{2008-Kosmrlj,2009-Kosmrlj}, but additionally find varying levels
of enrichment at different CDR3$\beta$ positions.
Furthermore, a large number of crystal structures of TCR\textendash peptide-MHC
complexes have been analyzed, that reveal a nonuniform contact interface between a TCR
and peptide-MHC (quantified, for example, by atom-atom contact profiles \textemdash
\ see Fig.~\ref{fig:Number-of-inter-atomic}), and show that positions 6 and 7 of the
CDR3$\beta$ sequence make the strongest contacts on average.
Taken together, these findings show that certain TCR positions are more important
than others for influencing thymic selection outcomes,
and that this information can be captured by nonuniform contact profiles such
as those in Fig.~\ref{fig:Number-of-inter-atomic}.

While Refs.~\cite{2008-Kosmrlj,2009-Kosmrlj} modeled thymic selection
outcomes depending on properties of inter-amino acid interactions,
they did not account for nonuniform contacts that a TCR makes with
peptide-MHC. 
In this paper, we develop a formalism to do so, and investigate how this
affects thymic selection outcomes. In particular, we consider two possible mechanisms
by which nonuniform contacts are mediated during thymic selection,
one of which we term \emph{deterministic}, and the other \emph{stochastic}.
We perform numerical simulations and analytical computations to characterize
the degree of enrichment of amino acids in the post-selection TCR
repertoire, and show that positions making stronger contacts end up
with greater degrees of enrichment.
While this may appear an expected outcome, the degree of enrichment depends
nontrivially on the entire contact profile, as well as the positive and negative
selection thresholds. In addition, we find that the interpretation of nonuniform contacts
remarkably affects the degree, and even sign, of enrichment.
Our study suggests a mechanistic origin for positional
differences in post-selection TCR amino acid enrichment that has been observed
in statistical analyses~\cite{2014-Callan-Mora-Walczak,2017-Callan-Mora-Walczak}
and experiments~\cite{2016-Stadinski},
and has implications for understanding how the functionality
of the post-selection repertoire emerges.

The paper is organized as follows:
in Section~\ref{sec:Model-description}, we develop a mathematical
model of thymic selection that incorporates nonuniform contact profiles.
In Section~\ref{sec:Nonuniform}, we consider two possible interpretations
of nonuniform contacts, and study their effects on thymic selection outcomes.
The stark differences between levels of enrichment from the two
interpretations are explained analytically. In Section~\ref{sec:Discussion},
we discuss further work and conclude.

\section{Model description\label{sec:Model-description}}

Immature T cells (or thymocytes) undergo positive and negative selection
in the thymus before maturation. Each thymocyte expresses a distinct
TCR on its surface, generated stochastically through V(D)J gene recombination
and insertions and deletions of nucleotides (whose probabilities have
been inferred from high-throughput sequencing data~\cite{2012-Mora-Walczak-Callan}),
creating a diverse pre-selection repertoire. In the thymic cortex,
thymocytes are presented with self-peptide-MHC by thymic antigen presenting cells (APCs).
Thymocytes that do not bind strongly enough to any self-peptide-MHC
die of insufficient survival signals; this is called positive selection.
Thymocytes that survive positive selection migrate to the thymic medulla,
where they are further screened against self-peptide-MHC, and those
that bind too strongly with at least one self-peptide-MHC receive
apoptotic signals and are eliminated; this is called negative selection.

\begin{figure*}
\subfloat[]{\includegraphics[scale=0.7]{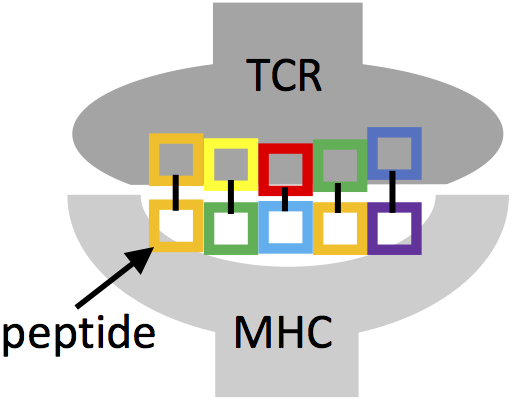}}\subfloat[]{\includegraphics[scale=0.45]{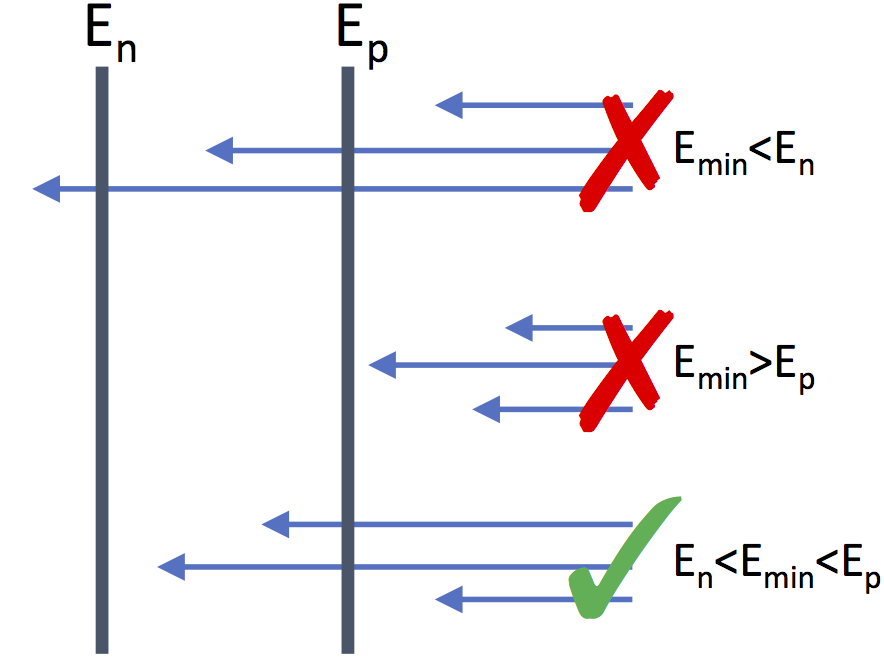}}\hfill{}\subfloat[]{\includegraphics[scale=0.5]{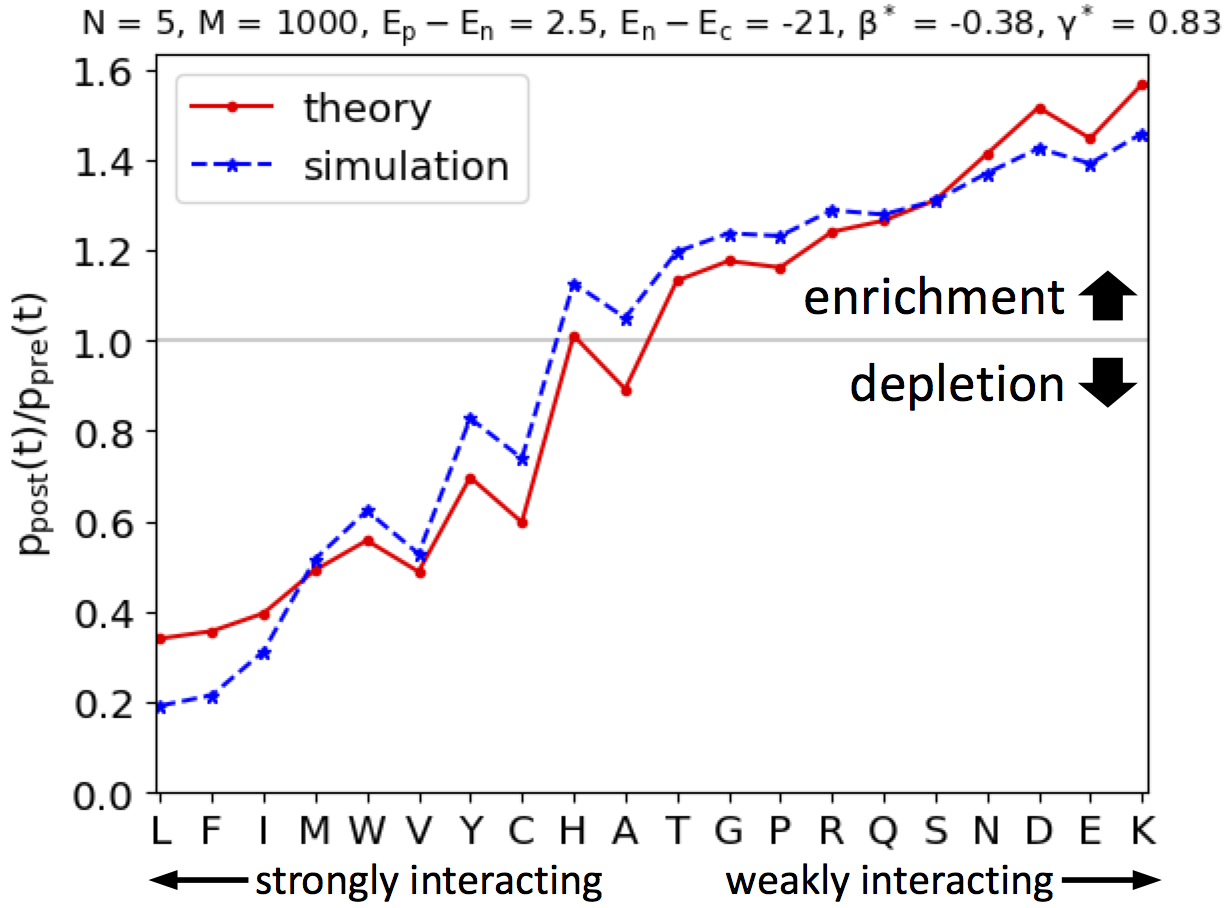}
}

\caption{Mathematical model of positive and negative selection in the thymus.
(a) String representation of TCR\textendash peptide-MHC binding. A
TCR sequence of length $N$ (here $N=5$) interacts with a peptide
sequence, also of length $N$, bound to a MHC molecule. Each colored
square represents a different amino acid. TCR and peptide amino acids
interact in a pairwise fashion (see Eq.~(\ref{eq:E(t,s)})). (b) Schematic
of possible thymic selection outcomes. Top: thymocyte failing negative
selection. Middle: thymocyte failing positive selection. Bottom: thymocyte
surviving positive and negative selection. (c) Enrichment curve showing
the amino acid distribution of TCR sequences surviving thymic selection,
divided by the pre-selection distribution $p_{\text{pre}}(t)$. Parameter
values (same as Ref.~\cite{2009-Kosmrlj}): all $f(c_{i})=1$, $N=5$,
$M=10^{3}$, $E_{p}-E_{n}=2.5$, $E_{n}-E_{c}=-21$ (in units of $k_{B}T$).
Blue dotted line: results of numerical simulations of the model for $5\times10^{6}$
independently generated TCR sequences, each encountering $M=10^{3}$
independent self-peptide sequences. Red solid line: prediction of
Eq.~(\ref{eq:factor_CaseAB}) with $\beta^{*}$ obtained from 
Eqs.~(\ref{eq:gamma_CaseAB})\textendash (\ref{eq:mu_G_CaseAB}).
Amino acids are arranged in order of
increasing $[J(t,a)]_{a}$ (see Eq.~(\ref{eq:mean_prior})). Enrichment
curves for all $i=1,\dots,N$ have been averaged together. A value
of 1 implies that the amino acid is observed equally often pre- and
post-selection; values greater (less) than 1 imply enrichment (depletion)
in the post-selection repertoire. Theory gave $\beta^{*}=-0.38$,
implying selection for weakly interacting amino acids (right side
of figure).\label{fig:model1}}
\end{figure*}

We cast this process in a mathematical model as follows: following
Ref.~\cite{2008-Kosmrlj}, TCR sequences, $\mathbf{t}=(t_{1},\dots,t_{N})$,
of length $N$ are generated by sampling amino acids $t_{i}$, $i=1,\dots,N$,
independently from a distribution $p_{\text{pre}}(t_{i})$, that is
taken to be the amino acid distribution of the human proteome~\cite{2017-Ensembl}.
While this is highly simplified compared to how the actual pre-selection
repertoire is generated~\cite{2012-Mora-Walczak-Callan}, it does
not affect our later results, which concern the action of thymic selection
on this distribution. Unlike Ref.~\cite{2008-Kosmrlj}, here a TCR
has the capacity to make nonuniform contacts, which we capture by
specifying $\mathbf{c}=(c_{1},\dots,c_{N})$. We describe values and
interpretations of $\mathbf{c}$ in the next section.

During thymic selection, a TCR interacts with $M$ independent self-peptide
sequences, $\mathbf{s}=(s_{1},\dots,s_{N})$, of length $N$ bound
to MHC, that are also randomly generated according to $p_{\text{pre}}$.
Because these are mostly linear peptides, they do not have an associated
$\mathbf{c}$. Following Ref.~\cite{2008-Kosmrlj}, we model the binding
strength between a TCR and self-peptide as pairwise interactions between
TCR amino acid $t_{i}$ and self-peptide amino acid $s_{i}$, for $i=1,\dots,N$
(see Fig.~\ref{fig:model1}a). Thus, the overall binding energy, $E(\mathbf{t},\mathbf{s},\mathbf{c})$,
is
\begin{equation}
E(\mathbf{t},\mathbf{s},\mathbf{c})=E_{c}+\sum_{i=1}^{N}f(c_{i})J(t_{i},s_{i}),\label{eq:E(t,s)}
\end{equation}
where $E_{c}$ captures interactions between TCR and MHC, $J(t,s)$
is an interaction potential that in principle captures biochemical
and other properties of inter-amino acid interactions, and $f(c_{i}$)
accounts for nonuniform contact at position $i$. Following Ref.~\cite{2008-Kosmrlj},
we use the Miyazawa\textendash Jernigan (MJ) matrix for $J(t,s)$~\cite{Miyazawa-Jernigan}, 
whose structure largely arises from hydrophobic
forces~\cite{2016-Stadinski,1997-Li-Tang-Wingreen}; hydrophobic amino acids are \emph{strongly}
interacting (more negative $J(t,s)$ values), while hydrophilic amino acids are
\emph{weakly} interacting (less negative $J(t,s)$ values).
(Note that the results that follow do not qualitatively depend on the potential,
but Ref.~\cite{2016-Stadinski} noted the importance of hydrophobicity for the strength of
TCR\textendash peptide-MHC interactions, implying that using the MJ matrix is reasonable.)
Reference~\cite{2008-Kosmrlj} implicitly assumed that all $f(c_{i})=1$. Henceforth, we will
specialize to one MHC type (i.e. $E_{c}$ a constant) because its diversity is
much lower than that of self-peptides (a human for example expresses
6 different MHC class I molecules), and we are not going to focus
on how TCR cross-reactivity to other MHC molecules can
arise~\cite{1999-Detours-Mehr-Perelson,1999-Detours-Perelson}.

Positive and negative selection are carried out as follows: if the
strongest (minimum) interaction energy between a TCR and $M$ independent
self-peptide-MHC is weaker (greater) than a positive selection threshold
$E_{p}$, or if it is stronger (less) than a negative selection threshold
$E_{n}$, the thymocyte dies. These hard constraints are consistent
with experiments that found relatively small differences in TCR\textendash ligand
affinity at the negative selection threshold~\cite{2006-Daniels}
(although they found that the positive selection threshold is softer).
Thymocytes that interact with $M$ self-peptide-MHC with strongest
(minimum) binding energy within $[E_{n},E_{p}]$ survive thymic selection
and mature into naïve T cells (see Fig.~\ref{fig:model1}b).

\paragraph{Parameter values:}

The CDR3 loops of a TCR typically make the greatest contacts with
peptide (as opposed to MHC)~\cite{2005-Garcia-Adams}. Here, the TCR
amino acid string $\mathbf{t}$ represents the CDR3$\beta$ sequence, which is typically
of length 10\textendash 18~\cite{2017-Callan-Mora-Walczak}.
During thymic selection, a thymocyte typically
interacts with $10^{3}\text{\textendash}10^{4}$ self-peptide-MHC.
Realistic values of the difference between positive and negative
selection thresholds, and the T cell activation free energy with self-peptide
alone (without MHC), are given by $E_{p}-E_{n}=2.5\,k_{B}T$ and $E_{n}-E_{c}=-21\,k_{B}T$,
respectively~\cite{2008-Kosmrlj,2009-Kosmrlj}.

\paragraph{Model without nonuniform contacts:}

References~\cite{2008-Kosmrlj,2009-Kosmrlj} considered a model with
all $f(c_{i})=1$, $N=5$, $M=10^{3}$, and $E_{p}-E_{n}$ and $E_{n}-E_{c}$
given above. Numerical simulations of this model resulted in a distribution
of TCR sequences surviving selection
that was statistically different
from the pre-selection one, in that the former was enriched in weakly
interacting amino acids and depleted in strongly interacting ones
(see Fig.~\ref{fig:model1}c).
This result was consistent with later experiments~\cite{2016-Stadinski}.

\subsection{Theory of the post-selection TCR repertoire distribution}

Reference~\cite{2009-Kosmrlj} developed an analytical theory for
the post-selection TCR repertoire distribution, valid in the limit of $N,M\to\infty$.
Here, we  extend this theory to include nonuniform contact
profiles. A self-contained derivation is in the Appendix; we state below
 the essential results.

A TCR with sequence $\mathbf{t}=(t_{1},\dots,t_{N})$ and contact
profile $\mathbf{c}=(c_{1},\dots,c_{N})$ experiences a distribution
of binding energies during thymic selection, with mean
\begin{equation}
\mu(\mathbf{t},\mathbf{c})=E_{c}+\sum_{i=1}^{N}\mu(t_{i},c_{i}),\label{eq:mean_t}
\end{equation}
and variance
\begin{equation}
\nu(\mathbf{t},\mathbf{c})=\sum_{i=1}^{N}\nu(t_{i},c_{i}),\label{eq:variance_t}
\end{equation}
where there are no cross-correlation terms in Eq.~(\ref{eq:variance_t}) because
sites are independent.
In Ref.~\cite{2009-Kosmrlj} (where all $f(c_{i})=1$), $\mu(t_{i},c_{i})$
and $\nu(t_{i},c_{i})$ are given by
\begin{equation}
\mu(t_{i})=[J(t_{i},a)]_{a},\label{eq:mean_t_i}
\end{equation}
and
\begin{equation}
\nu(t_{i})=[J(t_{i},a)^{2}]_{a}-[J(t_{i},a)]_{a}^{2},\label{eq:variance_t_i}
\end{equation}
where $[J(t,a)]_{a}$ and $[J(t,a)^2]_{a}$ are the first and second moments
of interaction strength experienced by amino acid $t$ when
interacting with self-peptide sequences, i.e.
\begin{equation}
[J(t,a)]_{a}\equiv\sum_{a=1}^{20}J(t,a)p_{\text{pre}}(a).\label{eq:mean_prior}
\end{equation}
In this paper, the forms of $\mu(t_{i},c_{i})$
and $\nu(t_{i},c_{i})$ depend on the interpretation of $\mathbf{c}$
and will be specified in the next section.

For large $N$ and $M$, the minimum binding energy experienced by
a TCR during selection tends to the Gumbel distribution (see Appendix),
whose peak is at
\begin{equation}
\rho_{G}(\mathbf{t},\mathbf{c})=\mu(\mathbf{t},\mathbf{c})-\sqrt{2\nu(\mathbf{t},\mathbf{c})}\alpha,\label{eq:a_M}
\end{equation}
where $\alpha=\sqrt{\log M}$, and whose variance is $\nu_{G}(\mathbf{t},\mathbf{c})=\pi^{2}\nu(\mathbf{t},\mathbf{c})/(12\log M)$.
In the limit of $N,M\to\infty$ (keeping $N\propto\log M$), this
distribution concentrates around its peak, which lies somewhere between
$E_{n}$ and $E_{p}$. In this limit, the post-selection TCR repertoire
distribution becomes (see Appendix)
\begin{equation}
p_{\text{post}}(t_{i}\big|c_{i})=\frac{1}{Z_{\beta,\gamma,c_{i}}}e^{-\beta[\mu(t_{i},c_{i})-\gamma\nu(t_{i},c_{i})]}p_{\text{pre}}(t_{i}),\label{eq:factor_CaseAB}
\end{equation}
where $Z_{\beta,\gamma,c_{i}}$ ensures normalization,
\begin{equation}
\gamma=\frac{\alpha}{\sqrt{2\sum_{i=1}^{N}\langle\nu(t_{i},c_{i})\rangle_{\beta,\gamma,c_{i}}}},\label{eq:gamma_CaseAB}
\end{equation}
and
\begin{equation}
\langle\nu(t_{i},c_{i})\rangle_{\beta,\gamma,c_{i}}\equiv\sum_{t_{i}=1}^{20}\nu(t_{i},c_{i})p_{\text{post}}(t_{i}\big|c_{i}).\label{eq:avgsel_CaseAB}
\end{equation}
The value of $\beta$ in Eq.~(\ref{eq:factor_CaseAB}) is chosen such that
$\langle\rho_{G}(\mathbf{t},\mathbf{c})\rangle$ (from Eq.~(\ref{eq:a_M}))
lies within $[E_{n},E_{p}]$. $p_{\text{post}}(t_{i}\big|c_{i})$
depends on $c_{i}$ explicitly, and on the full contact profile implicitly
through $\gamma$. Also in this limit, $\langle\rho_{G}(\mathbf{t},\mathbf{c})\rangle$
becomes
\begin{equation}
\langle\rho_{G}(\mathbf{t},\mathbf{c})\rangle_{\beta,\gamma,\mathbf{c}}=\sum_{i=1}^{N}\langle\mu(t_{i},c_{i})-\gamma\nu(t_{i},c_{i})\rangle_{\beta,\gamma,c_{i}}-\frac{\alpha^{2}}{2\gamma}.\label{eq:mu_G_CaseAB}
\end{equation}

In practice, for each value of $\beta$, one iterates between Eqs.
(\ref{eq:gamma_CaseAB}) and (\ref{eq:avgsel_CaseAB}) until a self-consistent
value of $\gamma$ is obtained. Values of $\beta$ corresponding to
$\langle\rho_{G}(\mathbf{t},\mathbf{c})\rangle_{\beta,\gamma,\mathbf{c}}=E_{p}$
and $\langle\rho_{G}(\mathbf{t},\mathbf{c})\rangle_{\beta,\gamma,\mathbf{c}}=E_{n}$
are found, and the one closer to zero is taken to be $\beta^{*}$
(or $\beta^{*}=0$ if they straddle 0). Intuitively, $\beta^{*}$
parametrizes the degree to which weakly or strongly interacting amino
acids are enriched in the post-selection TCR repertoire; positive
$\beta^{*}$ implies selecting for strongly interacting amino acids,
and negative $\beta^{*}$ implies selecting for weakly interacting
ones.


\section{Interpretations of nonuniform TCR contact profiles\label{sec:Nonuniform}}

\begin{figure}
\subfloat[]{\includegraphics[scale=0.6]{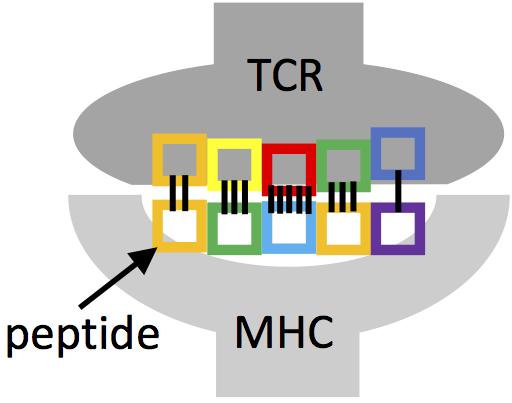}

}\subfloat[]{\includegraphics[scale=0.6]{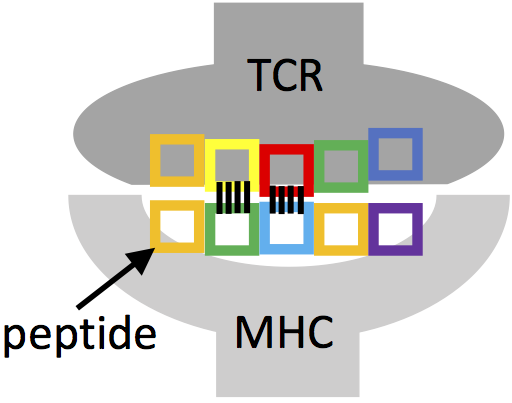}

}\caption{Schematic of deterministic (a) and stochastic (b) interpretations
of nonuniform TCR contact profiles. (a) Interactions between TCR amino
acid $t_{i}$ and peptide amino acid $s_{i}$ at position $i$ are
screened by a factor $c_{i}$, here represented as contacts of different
sizes. (b) Interactions between TCR and peptide amino acids at position
$i$ are made with probability $c_{i}$, $0\leq c_{i}\leq1$; this
is representated as contacts of the same size that are either present
or absent.\label{fig:Schematic-of-Cases}}
\end{figure}

How is the picture of Fig.~\ref{fig:model1}c modified when nonuniform
contact profiles are taken into account? We were initially inspired
by crystal structures of TCR\textendash peptide-MHC complexes, that
show a variation in the number of contacts made between TCR and peptide-MHC
along the TCR sequence for every structure examined (see Fig.~\ref{fig:Number-of-inter-atomic}); this contact profile need not even have
a single maximum. However, when many contact profiles were added
together, it appeared that positions 6 and 7 of the CDR3$\beta$ sequence
made the greatest contacts with peptide-MHC on average (see Fig.~\ref{fig:CaseAB}a),
which is consistent with experimental findings~\cite{2016-Stadinski} and
suggests that nonuniform contact profiles are useful in capturing positional
differences that influence thymic selection outcomes.
Figure~\ref{fig:CaseAB}a was obtained by rescaling the total number of contacts
measured from 53 TCR\textendash peptide-MHC structures
such that the maximum is 1. We will make use of this representative,
``average'' contact profile as $\mathbf{c}$ in the following.

\subsection{Deterministic screening interpretation}

Motivated by crystal structure analyses, we first considered a model
where every interaction between TCR amino acid
$t_{i}$ and self-peptide amino acid $s_{i}$ was weakened by a factor
$c_{i}$, $0\leq c_{i}\leq1$. This is represented by setting $f(c_{i})=c_{i}$
in Eq.~(\ref{eq:E(t,s)}). In principle, there is no reason to exclude
values of $c_{i}>1$
(depending on how interactions are modified by structure),
but here we only considered reduced interactions, for example due to screening by
intervening water molecules.

\begin{figure*}
\subfloat[]{\includegraphics[scale=0.4]{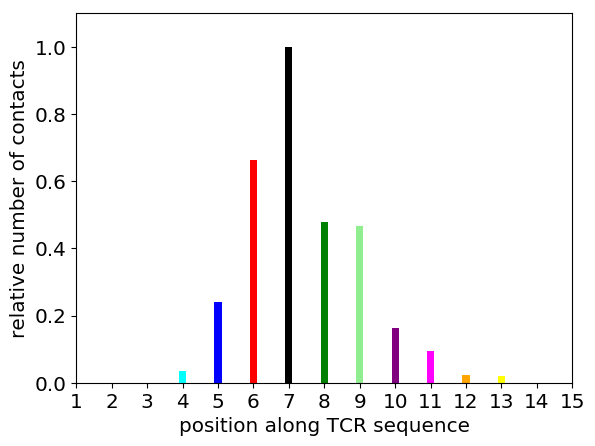}}

\subfloat[]{\includegraphics[scale=0.45]{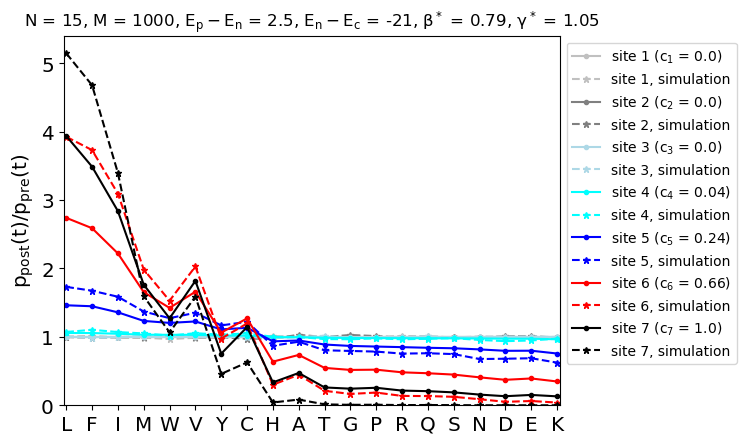}}\subfloat[]{\includegraphics[scale=0.45]{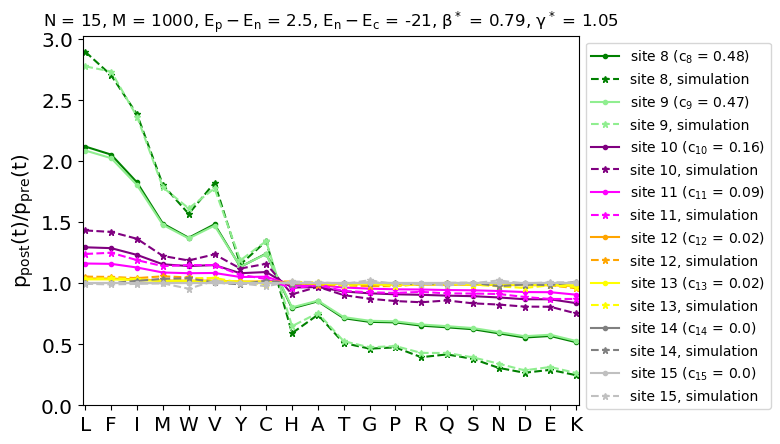}}

\subfloat[]{\includegraphics[scale=0.45]{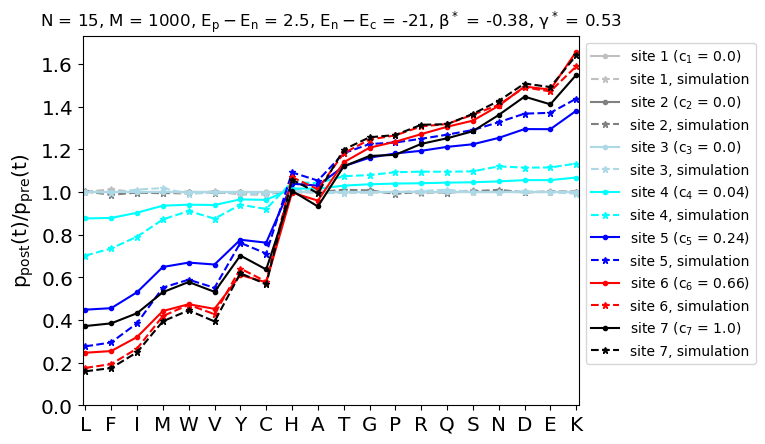}}\subfloat[]{\includegraphics[scale=0.45]{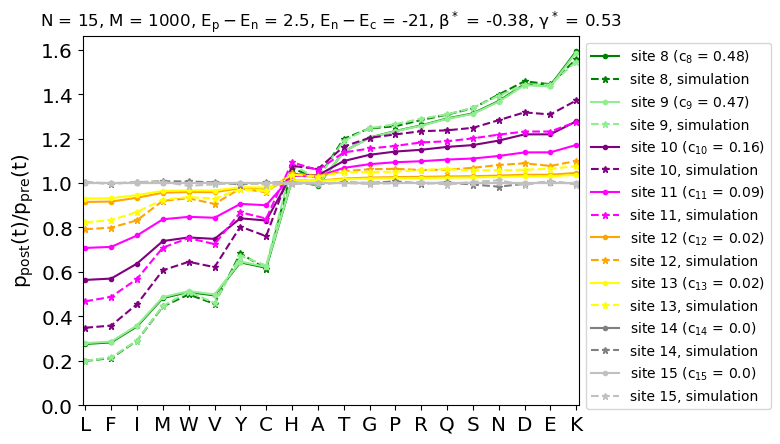}}

\caption{Thymic selection with nonuniform contact profiles. (a) Total number
of atom-atom contacts for 53 TCR\textendash peptide-MHC crystal structures,
rescaled such that the maximum is 1 (contact data from Ref.~\cite{2016-Stadinski};
see Protein Data Bank IDs therein). (b-e) Numerical simulations (dotted lines) and
analytical computations (solid lines) for the deterministic (b-c)
and stochastic (d-e) interpretations of nonuniform contacts. Each colored line represents
the degree of enrichment of amino acids at a position along
TCR sequences surviving thymic selection. For ease of viewing, sites 1\textendash 7
are plotted in (b,d), and sites 8\textendash 15 in (c,e). To create
the dotted lines, $5\times10^{6}$ simulations were performed for
each model with independently drawn TCR and self-peptide sequences.
Solid lines are plots of Eq.~(\ref{eq:factor_CaseAB}) with $\beta^{*}$
obtained from Eqs.~(\ref{eq:gamma_CaseAB})\textendash (\ref{eq:mu_G_CaseAB}).
Strongly interacting amino acids were enriched ($\beta^{*}>0$) for
the deterministic model, but depleted ($\beta^{*}<0$) for the stochastic
model. Parameter values: $N=15$, $M=10^{3}$, $E_{p}-E_{n}=2.5$,
$E_{n}-E_{c}=-21$ (in units of $k_{B}T$). Note: in (c) and (e),
the curves for sites 8 and 9 almost coincide as $c_{8}$ and $c_{9}$
are almost equal.\label{fig:CaseAB}}
\end{figure*}

We performed numerical simulations of our model using the deterministic interpretation
of the contact profile of Fig.~\ref{fig:CaseAB}a, with TCRs of length $N=15$ and keeping
the other parameter values the same as in Ref.~\cite{2009-Kosmrlj} and
Fig.~\ref{fig:model1}c (i.e. $M=10^{3}$, $E_{p}-E_{n}=2.5\,k_{B}T$, and $E_{n}-E_{c}=-21\,k_{B}T$).
Statistics of TCR amino acids in the post-selection repertoire are
shown in Figs.~\ref{fig:CaseAB}b\textendash c (dotted lines). Enrichment
curves for each site are plotted in a different color. Now, the degree
of enrichment depends on position; sites corresponding to the largest
$c_{i}$ (sites 6 and 7) experience the greatest degree of enrichment,
while sites making no contact (sites 1\textendash 3, 14 and 15) have
enrichment values close to 1. Furthermore, sites making contacts are
now enriched in \emph{strongly interacting} amino acids, which is
opposite of Fig.~\ref{fig:model1}c!

To understand this result, we repeated the previous theoretical analysis.
The mean and variance of interaction energies for amino acid $t_{i}$
at position $i$ are now modified to
\begin{align}
\mu_{A}(t_{i},c_{i}) & =c_{i}[J(t_{i},a)]_{a},\label{eq:mean_t_CaseA}
\end{align}
and
\begin{equation}
\nu_{A}(t_{i},c_{i})=c_{i}^{2}[J(t_{i},a)^{2}]_{a}-c_{i}^{2}[J(t_{i},a)]_{a}^{2},\label{eq:variance_t_CaseA}
\end{equation}
by replacing $J(t_{i},a)\to c_i J(t_{i},a)$ in Eqs.~(\ref{eq:mean_t_i}) and (\ref{eq:variance_t_i}).
Using these expressions in Eqs.~(\ref{eq:gamma_CaseAB})\textendash (\ref{eq:mu_G_CaseAB}),
we found $\beta^{*}=0.79>0$, which indeed implies enrichment of
strongly interacting amino acids. Plots of Eq.~(\ref{eq:factor_CaseAB})
for each position (solid lines in Figs.~\ref{fig:CaseAB}b\textendash c)
also resemble enrichment curves obtained from simulations.

Upon reflection, this change is not surprising.
The contact profile of Fig.~\ref{fig:CaseAB}a has an ``effective
length'' of $\sum_{i=1}^{15}c_{i}\approx3.2<5$, implying that the
mean binding energy experienced by a TCR during selection is roughly $64\%$
of that experienced in the model with $N=5$ and without nonuniform
contacts which produced Fig.~\ref{fig:model1}c (and indeed, we find $\beta^*>0$
for $N=3$ and without nonuniform contacts). Thus, it is plausible
that here, TCR sequences need to be enriched in more strongly interacting
amino acids in order to have their strongest binding energy during
selection fall between $E_{n}$ and $E_{p}$. Indeed, the analytical
theory makes this intuition concrete.

Note that the analytical curves systematically under-predict the
degree of enrichment, i.e. the value of $\beta^{*}$ obtained from theory was
slightly too small. For finite $M$, the extreme value distribution
in fact has a finite width, and so one might underestimate $\beta^{*}$
by matching $\langle\rho_{G}(\mathbf{t},\mathbf{c})\rangle_{\beta,\gamma,\mathbf{c}}$
to $E_{p}$. Indeed, when we include the next-order correction to
$\rho_{G}(\mathbf{t},\mathbf{c})$, which is positive (see after Eq.~(\ref{eq:a_M-1})
in the Appendix), the analytical curves match those from simulations
much better (not shown).

However, these results contradict the experiments of Ref.~\cite{2016-Stadinski},
as TCRs with strongly interacting amino acids at sites 6 and 7 should
fail negative selection. This implies that some of our assumptions,
for example that the interaction potential diminishes proportionally
with the contact profile (i.e. $f(c_{i})\propto c_{i}$), are incorrect.

\subsection{Stochastic binding interpretation}

In fact, crystal structures are merely static pictures of TCR\textendash peptide-MHC
binding, whereas the events leading to T cell activation following encounters
 with peptide-MHC are much more dynamic~\cite{2005-Garcia-Adams}.
Indeed, it is known that the CDR3$\beta$ loop of TCRs is relatively
flexible, with a range of conformations that can bind different ligands~\cite{2012-Baker}. 
There also exist crystal structures of the same
TCR bound to different peptide-MHC that show different parts of the
TCR binding the different ligands~\cite{2007-Colf}.

These facts motivate another interpretation of nonuniform contact profiles:
for every encounter during thymic selection, TCR amino acid $t_{i}$
binds to self-peptide amino acid $s_{i}$ with probability $c_{i}$,
$0\leq c_{i}\leq1$. This is represented by setting $f(c_{i})=X_{i}$, where $X_{i}$ is a Bernoulli
random variable with parameter $c_{i}$. This \emph{stochastic binding}
interpretation gives the same average interaction energy as the deterministic
one:
\begin{align}
\mu_{B}(t_{i},c_{i}) & =c_{i}[J(t_{i},a)]_{a},\label{eq:mean_t_CaseB}
\end{align}
while its variance is instead
\begin{align}
\nu_{B}(t_{i},c_{i}) & =c_{i}[J(t_{i},a)^{2}]_{a}-c_{i}^{2}[J(t_{i},a)]_{a}^{2},\label{eq:variance_t_CaseB}
\end{align}
which has one factor of $c_{i}$ multiplying $[J(t_{i},a)^{2}]_{a}$
(as opposed to two in Eq.~(\ref{eq:variance_t_CaseA})).

We performed numerical simulations of this model, and obtained a post-selection
repertoire shown in Figs.~\ref{fig:CaseAB}d\textendash e (dotted
lines). Again, the degree of enrichment is greater for sites making greater
contacts. However, unlike the deterministic interpretation, here sites making contacts
are enriched in \emph{weakly interacting} amino acids. Using Eqs.~(\ref{eq:mean_t_CaseB})
and (\ref{eq:variance_t_CaseB}), the analytical computations of
 Eqs.~(\ref{eq:gamma_CaseAB})\textendash(\ref{eq:mu_G_CaseAB})
also give $\beta^{*}=-0.38<0$. Thus, the argument that we made before was incomplete;
even though the ``effective length'' here is approximately 3.2 as well
because the mean interaction energies are equal (Eqs.~(\ref{eq:mean_t_CaseA}) vs.~(\ref{eq:mean_t_CaseB})),
their variances are different, and it is the \emph{variance}
that nontrivially modifies thymic selection outcomes!
Because of the larger variance in this case, it is plausible that weakly interacting
amino acids are sufficient for having the strongest of $M$ binding energies reach below $E_p$.
Again, the analytical theory makes this intuition concrete.

Another way to think of this is in terms of extreme values in \emph{lengths}.
During each interaction, a subset of positions of a TCR interact with
self-peptide amino acids. In the negative selection-dominated regime,
TCRs are more likely to be eliminated because of one interaction
that is too strong, rather than all $M$ interactions being too weak.
Using the analytical theory for $N\geq 7$ and without nonuniform contacts
gives $\beta^*=-\infty$, implying that no TCR that makes 7 or more contacts
during selection should survive. Also, the analytical theory for $N=6$ and without
nonuniform contacts gives $\beta^*=-1.7$, implying that
TCRs making 6 contacts during selection have a high probability of failing negative selection
too. Thus, the majority of TCRs that survive selection will have made
their strongest interaction with $\sim$5 contacts,
and indeed, the enrichment curves for sites 6\textendash 9
resemble that of Fig.~\ref{fig:model1}c.

The predicted curves from Eq.~(\ref{eq:factor_CaseAB}) (solid lines
in Figs.~\ref{fig:CaseAB}d\textendash e) agree well with numerical
results. Note that the theory predicts that site 7 is less enriched
than site 6, even through $c_{7}>c_{6}$ (compare black and red solid curves
in Fig.~\ref{fig:CaseAB}d)! That the degree of enrichment is not
monotonically related to $c_{i}$ can be seen by differentiating
$\mu_{B}(t_{i},c_{i})-\gamma\nu_{B}(t_{i},c_{i})$ w.r.t. $c_{i}$,
which reveals that according to this theory, there is an optimum $0<c_{i}<1$
that gives the greatest enrichment, which is different for each amino acid (because
this depends on $[J(t,a)]_{a}$ and $[J(t,a)^{2}]_{a}$).
(Repeating this calculation for the deterministic model however gives an optimal
$c_i$ that is negative, implying that there is monotonicity for that case.)
Numerical simulations however do not show this non-monotonicity,
and so we believe this is a manifestation of finite $M$ and $N$.

To summarize, we have shown how two possible interpretations
of nonuniform TCR contacts modify thymic selection outcomes in different ways.
We have also explained our results using an analytical theory valid in the
limit of large $N$ and $M\propto\ln N$.

\section{Discussion\label{sec:Discussion}}

In this paper, we presented a formalism to incorporate information
about TCR structure, through its nonuniform contact profile, into
a model of thymic selection. We showed how this leaves statistical signatures
at different positions of post-selection TCR sequences.
Importantly, we showed how these signatures depend on implementation of
nonuniform contacts, as a deterministic screening of interactions, or as probabilities
of stochastic binding events. In the actual thymus, these and other
scenarios probably play a role, and it would be interesting to
quantify their relative roles.

While we have added a further degree of realism to modeling TCR\textendash peptide-MHC
interactions, many features have been left out, such as:
\begin{itemize}
\item We did not account for nonuniform contacts with MHC. This is important
because a TCR binding more strongly with MHC might require weaker
interactions with peptide in order to be activated; this has been
studied previously to explain TCR cross-reactivity with foreign MHC
molecules~\cite{1999-Detours-Mehr-Perelson,1999-Detours-Perelson}.
Mathematically, this may be included into the model by modulating
$E_{c}$ by $f(c_c)$.
\item Stochastic binding events may be correlated between neighboring TCR
positions. Mathematically, this introduces cross-correlation terms
into the variance of interaction energies (Eq.~(\ref{eq:variance_t_CaseB})),
and makes the Legendre transform and self-averaging within the theory
more complicated (see Appendix).
\item The pre-selection TCR distribution resulting from V(D)J recombination
is not factorizable into $\prod_{i=1}^{N}p_{\text{pre}}(t_{i})$,
which also introduces correlations into $P_{\text{post}}(\mathbf{t})$.
\item The same TCR amino acid may contact more than one peptide-MHC amino
acid, and the same peptide-MHC amino acid may contact more than one
TCR amino acid.
\end{itemize}

Nonuniform contact profiles are a step towards modeling the complicated,
structure-dependent nature of TCR\textendash peptide-MHC interactions.
However, a more informed method should be developed to infer such a
profile from crystal structures. For example, we made use of measurements of
the number of peptide-MHC atoms a distance of 4Å away from TCR amino acids, but characteristic
distances should depend on the kind of interaction (hydrogen bonding, van der Waals, etc.).

We limited computations in this paper to one contact profile
to illustrate its effect on thymic selection outcomes. Separately,
it would be interesting to characterize the statistics of contact profiles from crystal structures.
This task however is limited by the relatively small number of crystal
structures known, as opposed to the large quantities of high-throughput sequencing
data available. We note that an additional step needs to be taken to connect the
model in this paper to statistics of aligned TCR sequences that appear in, 
e.g., Refs.~\cite{2014-Callan-Mora-Walczak,2017-Callan-Mora-Walczak}
(which do not focus on peptide contact residues but rather the entire aligned CDR3 region,
and hence do not find enrichment at TCR positions corresponding to recent
experiments~\cite{2016-Stadinski}): the model should be run separately for
different contact profiles, and the enrichment curves should be averaged
together according to the \emph{passing rates} for the different contact profiles.
Note that this is a possible mechanism for obtaining enrichment of an amino acid at one position
in a sequence alignment, and depletion of the \emph{same} amino acid at another,
because these positions might feature in \emph{different} contact profiles that
have different $\beta^*$. This also implies that the results from an \emph{average} contact
profile are, in general, different from running the model for separate contact profiles
and averaging the results together, because the former does not account for different
passing rates. Thus, it would be interesting to attempt the inverse problem of inferring
differential contacts and binding tendencies at different positions of a TCR sequence from
positional differences in the post-selection TCR repertoire, but this is complicated
by the under-determinacy of the problem.

Inferring overall patterns determining a TCR's specificity to peptide-MHC
from knowledge of TCR\textendash peptide-MHC crystal structures
is challenging because there is no one canonical way by which
a TCR interacts with peptide-MHC~\cite{2005-Garcia-Adams,1998-Ysern-Li-Mariuzza,2003-Housset-Malissen};
the same TCR may bind different peptide-MHC in very different ways
\cite{2007-Colf}. Predicting TCR sequences that recognize
a given set of peptide-MHC by inferring ``sequence motifs'' of TCRs has been
achieved very recently~\cite{2017-Dash,2017-Davis}. We believe that
analyzing thymic selection outcomes has implications for antigenic
specificity, because surviving thymic selection involves interactions
with a large number of self-peptide-MHC, and thus features relevant for
thymic selection outcomes are also relevant for antigenic
specificity. The features we have studied here have measurable effects
on the post-selection repertoire, and thus they probably contribute to antigenic
specificity as well, which perhaps gives a mechanistic basis for the
sequence motifs discovered in the recent studies~\cite{2017-Dash,2017-Davis}.

Recently, a paper that also modified the thymic selection
model  of Refs.~\cite{2008-Kosmrlj,2009-Kosmrlj} to include positional
differences in TCR\textendash peptide-MHC interactions appeared~\cite{2017-George}.
In essence, for a given TCR, they drew the values of $f(c_{i})J(t_{i},s_{i})$
in Eq.~(\ref{eq:E(t,s)}), $i=1,\dots,N$, independently from a Gaussian
distribution. Thus, they reduced thymic selection to an extreme value
problem with a random energy model, in which ``TCRs'' and ``amino acids''
lose their meaning; this contradicts studies that find predictive features determining specificity
that are based on TCR amino acid sequences~\cite{2016-Stadinski,2017-Dash,2017-Davis}.
However, their model is simpler to analyze and may be a useful null model,
and it would be interesting to compare it with the results from averaging
many different contact profiles together. The authors also commented that the
model of Refs.~\cite{2008-Kosmrlj,2009-Kosmrlj} fell short in that
very few self-peptides (i.e. the most strongly interacting ones) perform
the job of negative selection equally effectively as the full panel
of $M=10^{3}\text{\textendash}10^{4}$ self-peptides. While this is
true (which follows directly from specifying an inter-amino acid interaction
potential), it is not known how large this fraction is for the real
thymic selection process. Experimentally, this could be tested by
engineering the thymus to contain peptides consisting of only strongly
interacting amino acids~\cite{2005-Huseby}. Also, it is possible that
these peptides are somehow found rarely or not at all in the
thymus, because self-peptides are chopped-up versions of actual proteins.
Large values of $M$ are likely still required to randomly
generate such special peptides.
We note that the model we study here moves away from the limitations they raised,
as different TCRs with different contact profiles need not bind
equally strongly with the same, strongly interacting peptide.

\begin{acknowledgments}
This work was supported by the Ragon Institute of MGH, MIT and Harvard
and an A{*}STAR Scholarship (to H.C.). M.K. acknowledges support from
NSF through grant number DMR-1708280.
\end{acknowledgments}

\bibliography{TCRnonuniform}

\providecommand{\noopsort}[1]{}\providecommand{\singleletter}[1]{#1}%
\begin{thebibliography}{25}%
\makeatletter
\providecommand \@ifxundefined [1]{%
 \@ifx{#1\undefined}
}%
\providecommand \@ifnum [1]{%
 \ifnum #1\expandafter \@firstoftwo
 \else \expandafter \@secondoftwo
 \fi
}%
\providecommand \@ifx [1]{%
 \ifx #1\expandafter \@firstoftwo
 \else \expandafter \@secondoftwo
 \fi
}%
\providecommand \natexlab [1]{#1}%
\providecommand \enquote  [1]{``#1''}%
\providecommand \bibnamefont  [1]{#1}%
\providecommand \bibfnamefont [1]{#1}%
\providecommand \citenamefont [1]{#1}%
\providecommand \href@noop [0]{\@secondoftwo}%
\providecommand \href [0]{\begingroup \@sanitize@url \@href}%
\providecommand \@href[1]{\@@startlink{#1}\@@href}%
\providecommand \@@href[1]{\endgroup#1\@@endlink}%
\providecommand \@sanitize@url [0]{\catcode `\\12\catcode `\$12\catcode
  `\&12\catcode `\#12\catcode `\^12\catcode `\_12\catcode `\%12\relax}%
\providecommand \@@startlink[1]{}%
\providecommand \@@endlink[0]{}%
\providecommand \url  [0]{\begingroup\@sanitize@url \@url }%
\providecommand \@url [1]{\endgroup\@href {#1}{\urlprefix }}%
\providecommand \urlprefix  [0]{URL }%
\providecommand \Eprint [0]{\href }%
\providecommand \doibase [0]{http://dx.doi.org/}%
\providecommand \selectlanguage [0]{\@gobble}%
\providecommand \bibinfo  [0]{\@secondoftwo}%
\providecommand \bibfield  [0]{\@secondoftwo}%
\providecommand \translation [1]{[#1]}%
\providecommand \BibitemOpen [0]{}%
\providecommand \bibitemStop [0]{}%
\providecommand \bibitemNoStop [0]{.\EOS\space}%
\providecommand \EOS [0]{\spacefactor3000\relax}%
\providecommand \BibitemShut  [1]{\csname bibitem#1\endcsname}%
\let\auto@bib@innerbib\@empty
\bibitem [{\citenamefont {Alam}\ \emph {et~al.}(1996)\citenamefont {Alam},
  \citenamefont {Travers}, \citenamefont {Wung}, \citenamefont {Nasholds},
  \citenamefont {Redpath}, \citenamefont {Jameson},\ and\ \citenamefont
  {Gascoigne}}]{1996-Alam}%
  \BibitemOpen
  \bibfield  {author} {\bibinfo {author} {\bibfnamefont {S.~M.}\ \bibnamefont
  {Alam}}, \bibinfo {author} {\bibfnamefont {P.~J.}\ \bibnamefont {Travers}},
  \bibinfo {author} {\bibfnamefont {J.~L.}\ \bibnamefont {Wung}}, \bibinfo
  {author} {\bibfnamefont {W.}~\bibnamefont {Nasholds}}, \bibinfo {author}
  {\bibfnamefont {S.}~\bibnamefont {Redpath}}, \bibinfo {author} {\bibfnamefont
  {S.~C.}\ \bibnamefont {Jameson}}, \ and\ \bibinfo {author} {\bibfnamefont
  {N.~R.~J.}\ \bibnamefont {Gascoigne}},\ }\href@noop {} {\bibfield  {journal}
  {\bibinfo  {journal} {Nature}\ }\textbf {\bibinfo {volume} {381}},\ \bibinfo
  {pages} {616} (\bibinfo {year} {1996})}\BibitemShut {NoStop}%
\bibitem [{\citenamefont {Krogsgaard}\ and\ \citenamefont
  {Davis}(2005)}]{2005-Krogsgaard-Davis}%
  \BibitemOpen
  \bibfield  {author} {\bibinfo {author} {\bibfnamefont {M.}~\bibnamefont
  {Krogsgaard}}\ and\ \bibinfo {author} {\bibfnamefont {M.~M.}\ \bibnamefont
  {Davis}},\ }\href@noop {} {\bibfield  {journal} {\bibinfo  {journal} {Nat.
  Immunol.}\ }\textbf {\bibinfo {volume} {6}},\ \bibinfo {pages} {239}
  (\bibinfo {year} {2005})}\BibitemShut {NoStop}%
\bibitem [{\citenamefont {Murugan}\ \emph {et~al.}(2012)\citenamefont
  {Murugan}, \citenamefont {Mora}, \citenamefont {Walczak},\ and\ \citenamefont
  {{C. G. Callan, Jr.}}}]{2012-Mora-Walczak-Callan}%
  \BibitemOpen
  \bibfield  {author} {\bibinfo {author} {\bibfnamefont {A.}~\bibnamefont
  {Murugan}}, \bibinfo {author} {\bibfnamefont {T.}~\bibnamefont {Mora}},
  \bibinfo {author} {\bibfnamefont {A.~M.}\ \bibnamefont {Walczak}}, \ and\
  \bibinfo {author} {\bibnamefont {{C. G. Callan, Jr.}}},\ }\href@noop {}
  {\bibfield  {journal} {\bibinfo  {journal} {Proc. Natl. Acad. Sci. U.S.A.}\
  }\textbf {\bibinfo {volume} {109}},\ \bibinfo {pages} {16161} (\bibinfo
  {year} {2012})}\BibitemShut {NoStop}%
\bibitem [{\citenamefont {Huseby}\ \emph {et~al.}(2005)\citenamefont {Huseby},
  \citenamefont {White}, \citenamefont {Crawford}, \citenamefont {Vass},
  \citenamefont {Becker}, \citenamefont {Pinilla}, \citenamefont {Marrack},\
  and\ \citenamefont {Kappler}}]{2005-Huseby}%
  \BibitemOpen
  \bibfield  {author} {\bibinfo {author} {\bibfnamefont {E.~S.}\ \bibnamefont
  {Huseby}}, \bibinfo {author} {\bibfnamefont {J.}~\bibnamefont {White}},
  \bibinfo {author} {\bibfnamefont {F.}~\bibnamefont {Crawford}}, \bibinfo
  {author} {\bibfnamefont {T.}~\bibnamefont {Vass}}, \bibinfo {author}
  {\bibfnamefont {D.}~\bibnamefont {Becker}}, \bibinfo {author} {\bibfnamefont
  {C.}~\bibnamefont {Pinilla}}, \bibinfo {author} {\bibfnamefont
  {P.}~\bibnamefont {Marrack}}, \ and\ \bibinfo {author} {\bibfnamefont
  {J.~W.}\ \bibnamefont {Kappler}},\ }\href@noop {} {\bibfield  {journal}
  {\bibinfo  {journal} {Cell}\ }\textbf {\bibinfo {volume} {122}},\ \bibinfo
  {pages} {247} (\bibinfo {year} {2005})}\BibitemShut {NoStop}%
\bibitem [{\citenamefont {Elhanati}\ \emph {et~al.}(2014)\citenamefont
  {Elhanati}, \citenamefont {Murugan}, \citenamefont {{C. G. Callan, Jr.}},
  \citenamefont {Mora},\ and\ \citenamefont
  {Walczak}}]{2014-Callan-Mora-Walczak}%
  \BibitemOpen
  \bibfield  {author} {\bibinfo {author} {\bibfnamefont {Y.}~\bibnamefont
  {Elhanati}}, \bibinfo {author} {\bibfnamefont {A.}~\bibnamefont {Murugan}},
  \bibinfo {author} {\bibnamefont {{C. G. Callan, Jr.}}}, \bibinfo {author}
  {\bibfnamefont {T.}~\bibnamefont {Mora}}, \ and\ \bibinfo {author}
  {\bibfnamefont {A.~M.}\ \bibnamefont {Walczak}},\ }\href@noop {} {\bibfield
  {journal} {\bibinfo  {journal} {Proc. Natl. Acad. Sci. U.S.A.}\ }\textbf
  {\bibinfo {volume} {111}},\ \bibinfo {pages} {9875} (\bibinfo {year}
  {2014})}\BibitemShut {NoStop}%
\bibitem [{\citenamefont {Sethna}\ \emph {et~al.}(2017)\citenamefont {Sethna},
  \citenamefont {Elhanati}, \citenamefont {Dudgeon}, \citenamefont {{C. G.
  Callan, Jr.}}, \citenamefont {Levine}, \citenamefont {Mora},\ and\
  \citenamefont {Walczak}}]{2017-Callan-Mora-Walczak}%
  \BibitemOpen
  \bibfield  {author} {\bibinfo {author} {\bibfnamefont {Z.}~\bibnamefont
  {Sethna}}, \bibinfo {author} {\bibfnamefont {Y.}~\bibnamefont {Elhanati}},
  \bibinfo {author} {\bibfnamefont {C.~R.}\ \bibnamefont {Dudgeon}}, \bibinfo
  {author} {\bibnamefont {{C. G. Callan, Jr.}}}, \bibinfo {author}
  {\bibfnamefont {A.~J.}\ \bibnamefont {Levine}}, \bibinfo {author}
  {\bibfnamefont {T.}~\bibnamefont {Mora}}, \ and\ \bibinfo {author}
  {\bibfnamefont {A.~M.}\ \bibnamefont {Walczak}},\ }\href@noop {} {\bibfield
  {journal} {\bibinfo  {journal} {Proc. Natl. Acad. Sci. U.S.A.}\ }\textbf
  {\bibinfo {volume} {114}},\ \bibinfo {pages} {2253} (\bibinfo {year}
  {2017})}\BibitemShut {NoStop}%
\bibitem [{\citenamefont {Stadinski}\ \emph {et~al.}(2016)\citenamefont
  {Stadinski}, \citenamefont {Shekhar}, \citenamefont {G{\'o}mez-Touri{\~n}o},
  \citenamefont {Jung}, \citenamefont {Sasaki}, \citenamefont {Sewell},
  \citenamefont {Peakman}, \citenamefont {Chakraborty},\ and\ \citenamefont
  {Huseby}}]{2016-Stadinski}%
  \BibitemOpen
  \bibfield  {author} {\bibinfo {author} {\bibfnamefont {B.~D.}\ \bibnamefont
  {Stadinski}}, \bibinfo {author} {\bibfnamefont {K.}~\bibnamefont {Shekhar}},
  \bibinfo {author} {\bibfnamefont {I.}~\bibnamefont {G{\'o}mez-Touri{\~n}o}},
  \bibinfo {author} {\bibfnamefont {J.}~\bibnamefont {Jung}}, \bibinfo {author}
  {\bibfnamefont {K.}~\bibnamefont {Sasaki}}, \bibinfo {author} {\bibfnamefont
  {A.~K.}\ \bibnamefont {Sewell}}, \bibinfo {author} {\bibfnamefont
  {M.}~\bibnamefont {Peakman}}, \bibinfo {author} {\bibfnamefont {A.~K.}\
  \bibnamefont {Chakraborty}}, \ and\ \bibinfo {author} {\bibfnamefont {E.~S.}\
  \bibnamefont {Huseby}},\ }\href@noop {} {\bibfield  {journal} {\bibinfo
  {journal} {Nat. Immunol.}\ }\textbf {\bibinfo {volume} {17}},\ \bibinfo
  {pages} {946} (\bibinfo {year} {2016})}\BibitemShut {NoStop}%
\bibitem [{\citenamefont {Yates}(2014)}]{2014-Yates}%
  \BibitemOpen
  \bibfield  {author} {\bibinfo {author} {\bibfnamefont {A.~J.}\ \bibnamefont
  {Yates}},\ }\href@noop {} {\bibfield  {journal} {\bibinfo  {journal} {Front.
  Immunol.}\ }\textbf {\bibinfo {volume} {5}},\ \bibinfo {pages} {13} (\bibinfo
  {year} {2014})}\BibitemShut {NoStop}%
\bibitem [{\citenamefont {Detours}\ \emph {et~al.}(1999)\citenamefont
  {Detours}, \citenamefont {Mehr},\ and\ \citenamefont
  {Perelson}}]{1999-Detours-Mehr-Perelson}%
  \BibitemOpen
  \bibfield  {author} {\bibinfo {author} {\bibfnamefont {V.}~\bibnamefont
  {Detours}}, \bibinfo {author} {\bibfnamefont {R.}~\bibnamefont {Mehr}}, \
  and\ \bibinfo {author} {\bibfnamefont {A.~S.}\ \bibnamefont {Perelson}},\
  }\href@noop {} {\bibfield  {journal} {\bibinfo  {journal} {J. Theor. Biol.}\
  }\textbf {\bibinfo {volume} {200}},\ \bibinfo {pages} {389} (\bibinfo {year}
  {1999})}\BibitemShut {NoStop}%
\bibitem [{\citenamefont {Detours}\ and\ \citenamefont
  {Perelson}(1999)}]{1999-Detours-Perelson}%
  \BibitemOpen
  \bibfield  {author} {\bibinfo {author} {\bibfnamefont {V.}~\bibnamefont
  {Detours}}\ and\ \bibinfo {author} {\bibfnamefont {A.~S.}\ \bibnamefont
  {Perelson}},\ }\href@noop {} {\bibfield  {journal} {\bibinfo  {journal}
  {Proc. Natl. Acad. Sci. U.S.A.}\ }\textbf {\bibinfo {volume} {96}},\ \bibinfo
  {pages} {5153} (\bibinfo {year} {1999})}\BibitemShut {NoStop}%
\bibitem [{\citenamefont {Ko{\v s}mrlj}\ \emph {et~al.}(2008)\citenamefont
  {Ko{\v s}mrlj}, \citenamefont {Jha}, \citenamefont {Huseby}, \citenamefont
  {Kardar},\ and\ \citenamefont {Chakraborty}}]{2008-Kosmrlj}%
  \BibitemOpen
  \bibfield  {author} {\bibinfo {author} {\bibfnamefont {A.}~\bibnamefont
  {Ko{\v s}mrlj}}, \bibinfo {author} {\bibfnamefont {A.~K.}\ \bibnamefont
  {Jha}}, \bibinfo {author} {\bibfnamefont {E.~S.}\ \bibnamefont {Huseby}},
  \bibinfo {author} {\bibfnamefont {M.}~\bibnamefont {Kardar}}, \ and\ \bibinfo
  {author} {\bibfnamefont {A.~K.}\ \bibnamefont {Chakraborty}},\ }\href@noop {}
  {\bibfield  {journal} {\bibinfo  {journal} {Proc. Natl. Acad. Sci. U.S.A.}\
  }\textbf {\bibinfo {volume} {105}},\ \bibinfo {pages} {16671} (\bibinfo
  {year} {2008})}\BibitemShut {NoStop}%
\bibitem [{\citenamefont {Ko{\v s}mrlj}\ \emph {et~al.}(2009)\citenamefont
  {Ko{\v s}mrlj}, \citenamefont {Chakraborty}, \citenamefont {Kardar},\ and\
  \citenamefont {Shakhnovich}}]{2009-Kosmrlj}%
  \BibitemOpen
  \bibfield  {author} {\bibinfo {author} {\bibfnamefont {A.}~\bibnamefont
  {Ko{\v s}mrlj}}, \bibinfo {author} {\bibfnamefont {A.~K.}\ \bibnamefont
  {Chakraborty}}, \bibinfo {author} {\bibfnamefont {M.}~\bibnamefont {Kardar}},
  \ and\ \bibinfo {author} {\bibfnamefont {E.~I.}\ \bibnamefont
  {Shakhnovich}},\ }\href@noop {} {\bibfield  {journal} {\bibinfo  {journal}
  {Phys. Rev. Lett.}\ }\textbf {\bibinfo {volume} {103}},\ \bibinfo {pages}
  {068103} (\bibinfo {year} {2009})}\BibitemShut {NoStop}%
\bibitem [{\citenamefont {Miyazawa}\ and\ \citenamefont
  {Jernigan}(1996)}]{Miyazawa-Jernigan}%
  \BibitemOpen
  \bibfield  {author} {\bibinfo {author} {\bibfnamefont {S.}~\bibnamefont
  {Miyazawa}}\ and\ \bibinfo {author} {\bibfnamefont {R.~L.}\ \bibnamefont
  {Jernigan}},\ }\href@noop {} {\bibfield  {journal} {\bibinfo  {journal} {J.
  Mol. Biol.}\ }\textbf {\bibinfo {volume} {256}},\ \bibinfo {pages} {623}
  (\bibinfo {year} {1996})}\BibitemShut {NoStop}%
\bibitem [{\citenamefont {{B. L. Aken et al.}}(2017)}]{2017-Ensembl}%
  \BibitemOpen
  \bibfield  {author} {\bibinfo {author} {\bibnamefont {{B. L. Aken et al.}}},\
  }\href@noop {} {\bibfield  {journal} {\bibinfo  {journal} {Nucleic Acids
  Res.}\ }\textbf {\bibinfo {volume} {45}},\ \bibinfo {pages} {D635} (\bibinfo
  {year} {2017})}\BibitemShut {NoStop}%
\bibitem [{\citenamefont {Li}\ \emph {et~al.}(1997)\citenamefont {Li},
  \citenamefont {Tang},\ and\ \citenamefont
  {Wingreen}}]{1997-Li-Tang-Wingreen}%
  \BibitemOpen
  \bibfield  {author} {\bibinfo {author} {\bibfnamefont {H.}~\bibnamefont
  {Li}}, \bibinfo {author} {\bibfnamefont {C.}~\bibnamefont {Tang}}, \ and\
  \bibinfo {author} {\bibfnamefont {N.~S.}\ \bibnamefont {Wingreen}},\
  }\href@noop {} {\bibfield  {journal} {\bibinfo  {journal} {Phys. Rev. Lett.}\
  }\textbf {\bibinfo {volume} {79}},\ \bibinfo {pages} {765} (\bibinfo {year}
  {1997})}\BibitemShut {NoStop}%
\bibitem [{\citenamefont {Daniels}\ \emph {et~al.}(2006)\citenamefont
  {Daniels}, \citenamefont {Teixeiro}, \citenamefont {Gill}, \citenamefont
  {Hausmann}, \citenamefont {Roubaty}, \citenamefont {Holmberg}, \citenamefont
  {Werlen}, \citenamefont {Holl{\"a}nder}, \citenamefont {Gascoigne},\ and\
  \citenamefont {Palmer}}]{2006-Daniels}%
  \BibitemOpen
  \bibfield  {author} {\bibinfo {author} {\bibfnamefont {M.~A.}\ \bibnamefont
  {Daniels}}, \bibinfo {author} {\bibfnamefont {E.}~\bibnamefont {Teixeiro}},
  \bibinfo {author} {\bibfnamefont {J.}~\bibnamefont {Gill}}, \bibinfo {author}
  {\bibfnamefont {B.}~\bibnamefont {Hausmann}}, \bibinfo {author}
  {\bibfnamefont {D.}~\bibnamefont {Roubaty}}, \bibinfo {author} {\bibfnamefont
  {K.}~\bibnamefont {Holmberg}}, \bibinfo {author} {\bibfnamefont
  {G.}~\bibnamefont {Werlen}}, \bibinfo {author} {\bibfnamefont {G.~A.}\
  \bibnamefont {Holl{\"a}nder}}, \bibinfo {author} {\bibfnamefont {N.~R.~J.}\
  \bibnamefont {Gascoigne}}, \ and\ \bibinfo {author} {\bibfnamefont
  {E.}~\bibnamefont {Palmer}},\ }\href@noop {} {\bibfield  {journal} {\bibinfo
  {journal} {Nature}\ }\textbf {\bibinfo {volume} {444}},\ \bibinfo {pages}
  {724} (\bibinfo {year} {2006})}\BibitemShut {NoStop}%
\bibitem [{\citenamefont {Garcia}\ and\ \citenamefont
  {Adams}(2005)}]{2005-Garcia-Adams}%
  \BibitemOpen
  \bibfield  {author} {\bibinfo {author} {\bibfnamefont {K.~C.}\ \bibnamefont
  {Garcia}}\ and\ \bibinfo {author} {\bibfnamefont {E.~J.}\ \bibnamefont
  {Adams}},\ }\href@noop {} {\bibfield  {journal} {\bibinfo  {journal} {Cell}\
  }\textbf {\bibinfo {volume} {122}},\ \bibinfo {pages} {333} (\bibinfo {year}
  {2005})}\BibitemShut {NoStop}%
\bibitem [{\citenamefont {Baker}\ \emph {et~al.}(2012)\citenamefont {Baker},
  \citenamefont {Scott}, \citenamefont {Blevins},\ and\ \citenamefont
  {Hawse}}]{2012-Baker}%
  \BibitemOpen
  \bibfield  {author} {\bibinfo {author} {\bibfnamefont {B.~M.}\ \bibnamefont
  {Baker}}, \bibinfo {author} {\bibfnamefont {D.~R.}\ \bibnamefont {Scott}},
  \bibinfo {author} {\bibfnamefont {S.~J.}\ \bibnamefont {Blevins}}, \ and\
  \bibinfo {author} {\bibfnamefont {W.~F.}\ \bibnamefont {Hawse}},\ }\href@noop
  {} {\bibfield  {journal} {\bibinfo  {journal} {Immunol. Rev.}\ }\textbf
  {\bibinfo {volume} {250}},\ \bibinfo {pages} {10} (\bibinfo {year}
  {2012})}\BibitemShut {NoStop}%
\bibitem [{\citenamefont {Colf}\ \emph {et~al.}(2007)\citenamefont {Colf},
  \citenamefont {Bankovich}, \citenamefont {Hanick}, \citenamefont {Bowerman},
  \citenamefont {Jones}, \citenamefont {Kranz},\ and\ \citenamefont
  {Garcia}}]{2007-Colf}%
  \BibitemOpen
  \bibfield  {author} {\bibinfo {author} {\bibfnamefont {L.~A.}\ \bibnamefont
  {Colf}}, \bibinfo {author} {\bibfnamefont {A.~J.}\ \bibnamefont {Bankovich}},
  \bibinfo {author} {\bibfnamefont {N.~A.}\ \bibnamefont {Hanick}}, \bibinfo
  {author} {\bibfnamefont {N.~A.}\ \bibnamefont {Bowerman}}, \bibinfo {author}
  {\bibfnamefont {L.~L.}\ \bibnamefont {Jones}}, \bibinfo {author}
  {\bibfnamefont {D.~M.}\ \bibnamefont {Kranz}}, \ and\ \bibinfo {author}
  {\bibfnamefont {K.~C.}\ \bibnamefont {Garcia}},\ }\href@noop {} {\bibfield
  {journal} {\bibinfo  {journal} {Cell}\ }\textbf {\bibinfo {volume} {129}},\
  \bibinfo {pages} {135} (\bibinfo {year} {2007})}\BibitemShut {NoStop}%
\bibitem [{\citenamefont {Ysern}\ \emph {et~al.}(1998)\citenamefont {Ysern},
  \citenamefont {Li},\ and\ \citenamefont {Mariuzza}}]{1998-Ysern-Li-Mariuzza}%
  \BibitemOpen
  \bibfield  {author} {\bibinfo {author} {\bibfnamefont {X.}~\bibnamefont
  {Ysern}}, \bibinfo {author} {\bibfnamefont {H.}~\bibnamefont {Li}}, \ and\
  \bibinfo {author} {\bibfnamefont {R.~A.}\ \bibnamefont {Mariuzza}},\
  }\href@noop {} {\bibfield  {journal} {\bibinfo  {journal} {Nat. Struct.
  Biol.}\ }\textbf {\bibinfo {volume} {5}},\ \bibinfo {pages} {412} (\bibinfo
  {year} {1998})}\BibitemShut {NoStop}%
\bibitem [{\citenamefont {Housset}\ and\ \citenamefont
  {Malissen}(2003)}]{2003-Housset-Malissen}%
  \BibitemOpen
  \bibfield  {author} {\bibinfo {author} {\bibfnamefont {D.}~\bibnamefont
  {Housset}}\ and\ \bibinfo {author} {\bibfnamefont {B.}~\bibnamefont
  {Malissen}},\ }\href@noop {} {\bibfield  {journal} {\bibinfo  {journal}
  {Trends Immunol.}\ }\textbf {\bibinfo {volume} {24}},\ \bibinfo {pages} {429}
  (\bibinfo {year} {2003})}\BibitemShut {NoStop}%
\bibitem [{\citenamefont {Dash}\ \emph {et~al.}(2017)\citenamefont {Dash},
  \citenamefont {Fiore-Gartland}, \citenamefont {Hertz}, \citenamefont {Wang},
  \citenamefont {Sharma}, \citenamefont {Souquette}, \citenamefont {Crawford},
  \citenamefont {Clemens}, \citenamefont {Nguyen}, \citenamefont {Kedzierska},
  \citenamefont {{N. L. La Gruta}}, \citenamefont {Bradley},\ and\
  \citenamefont {Thomas}}]{2017-Dash}%
  \BibitemOpen
  \bibfield  {author} {\bibinfo {author} {\bibfnamefont {P.}~\bibnamefont
  {Dash}}, \bibinfo {author} {\bibfnamefont {A.~J.}\ \bibnamefont
  {Fiore-Gartland}}, \bibinfo {author} {\bibfnamefont {T.}~\bibnamefont
  {Hertz}}, \bibinfo {author} {\bibfnamefont {G.~C.}\ \bibnamefont {Wang}},
  \bibinfo {author} {\bibfnamefont {S.}~\bibnamefont {Sharma}}, \bibinfo
  {author} {\bibfnamefont {A.}~\bibnamefont {Souquette}}, \bibinfo {author}
  {\bibfnamefont {J.~C.}\ \bibnamefont {Crawford}}, \bibinfo {author}
  {\bibfnamefont {E.~B.}\ \bibnamefont {Clemens}}, \bibinfo {author}
  {\bibfnamefont {T.~H.~O.}\ \bibnamefont {Nguyen}}, \bibinfo {author}
  {\bibfnamefont {K.}~\bibnamefont {Kedzierska}}, \bibinfo {author}
  {\bibnamefont {{N. L. La Gruta}}}, \bibinfo {author} {\bibfnamefont
  {P.}~\bibnamefont {Bradley}}, \ and\ \bibinfo {author} {\bibfnamefont
  {P.~G.}\ \bibnamefont {Thomas}},\ }\href@noop {} {\bibfield  {journal}
  {\bibinfo  {journal} {Nature}\ }\textbf {\bibinfo {volume} {547}},\ \bibinfo
  {pages} {89} (\bibinfo {year} {2017})}\BibitemShut {NoStop}%
\bibitem [{\citenamefont {Glanville}\ \emph {et~al.}(2017)\citenamefont
  {Glanville}, \citenamefont {Huang}, \citenamefont {Nau}, \citenamefont
  {Hatton}, \citenamefont {Wagar}, \citenamefont {Rubelt}, \citenamefont {Ji},
  \citenamefont {Han}, \citenamefont {Krams}, \citenamefont {Pettus},
  \citenamefont {Haas}, \citenamefont {{C. S. Lindestam Arlehamn}},
  \citenamefont {Sette}, \citenamefont {Boyd}, \citenamefont {Scriba},
  \citenamefont {Martinez},\ and\ \citenamefont {Davis}}]{2017-Davis}%
  \BibitemOpen
  \bibfield  {author} {\bibinfo {author} {\bibfnamefont {J.}~\bibnamefont
  {Glanville}}, \bibinfo {author} {\bibfnamefont {H.}~\bibnamefont {Huang}},
  \bibinfo {author} {\bibfnamefont {A.}~\bibnamefont {Nau}}, \bibinfo {author}
  {\bibfnamefont {O.}~\bibnamefont {Hatton}}, \bibinfo {author} {\bibfnamefont
  {L.~E.}\ \bibnamefont {Wagar}}, \bibinfo {author} {\bibfnamefont
  {F.}~\bibnamefont {Rubelt}}, \bibinfo {author} {\bibfnamefont
  {X.}~\bibnamefont {Ji}}, \bibinfo {author} {\bibfnamefont {A.}~\bibnamefont
  {Han}}, \bibinfo {author} {\bibfnamefont {S.~M.}\ \bibnamefont {Krams}},
  \bibinfo {author} {\bibfnamefont {C.}~\bibnamefont {Pettus}}, \bibinfo
  {author} {\bibfnamefont {N.}~\bibnamefont {Haas}}, \bibinfo {author}
  {\bibnamefont {{C. S. Lindestam Arlehamn}}}, \bibinfo {author} {\bibfnamefont
  {A.}~\bibnamefont {Sette}}, \bibinfo {author} {\bibfnamefont {S.~D.}\
  \bibnamefont {Boyd}}, \bibinfo {author} {\bibfnamefont {T.~J.}\ \bibnamefont
  {Scriba}}, \bibinfo {author} {\bibfnamefont {O.~M.}\ \bibnamefont
  {Martinez}}, \ and\ \bibinfo {author} {\bibfnamefont {M.~M.}\ \bibnamefont
  {Davis}},\ }\href@noop {} {\bibfield  {journal} {\bibinfo  {journal}
  {Nature}\ }\textbf {\bibinfo {volume} {547}},\ \bibinfo {pages} {94}
  (\bibinfo {year} {2017})}\BibitemShut {NoStop}%
\bibitem [{\citenamefont {George}\ \emph {et~al.}(2017)\citenamefont {George},
  \citenamefont {Kessler},\ and\ \citenamefont {Levine}}]{2017-George}%
  \BibitemOpen
  \bibfield  {author} {\bibinfo {author} {\bibfnamefont {J.~T.}\ \bibnamefont
  {George}}, \bibinfo {author} {\bibfnamefont {D.~A.}\ \bibnamefont {Kessler}},
  \ and\ \bibinfo {author} {\bibfnamefont {H.}~\bibnamefont {Levine}},\
  }\href@noop {} {\bibfield  {journal} {\bibinfo  {journal} {Proc. Natl. Acad.
  Sci. U.S.A.}\ }\textbf {\bibinfo {volume} {114}},\ \bibinfo {pages} {E7875}
  (\bibinfo {year} {2017})}\BibitemShut {NoStop}%
\bibitem [{\citenamefont {Kardar}(1983)}]{1983-Kardar}%
  \BibitemOpen
  \bibfield  {author} {\bibinfo {author} {\bibfnamefont {M.}~\bibnamefont
  {Kardar}},\ }\href@noop {} {\bibfield  {journal} {\bibinfo  {journal} {Phys.
  Rev. Lett.}\ }\textbf {\bibinfo {volume} {51}},\ \bibinfo {pages} {523}
  (\bibinfo {year} {1983})}\BibitemShut {NoStop}%
\end{thebibliography}%

\appendix*
\section{Derivation of  the post-selection TCR repertoire
distribution for large $N$, $M$\label{sec:Theory}}

Here, we provide a self-contained derivation of a theory for the post-selection
TCR repertoire distribution, extending Ref.~\cite{2009-Kosmrlj} to
capture a nonuniform contact profile. The probability that a TCR with
sequence $\mathbf{t}$ and contact profile $\mathbf{c}$ survives
selection, $P(\text{post}\big|\mathbf{t},\mathbf{c})$, is equal to
the probability that the minimum of $M$ binding energies it encountered
lies within $[E_{n},E_{p}]$. Now, the binding energy $E(\mathbf{t},\mathbf{s},\mathbf{c})$
between a TCR and self-peptide sequence $\mathbf{s}$ bound to MHC
will be Gaussian distributed for large $N$, by the Central Limit
Theorem (because Eq.~(\ref{eq:E(t,s)}) contains a sum of $N$ independent,
but not identically distributed, $f(c_{i})J(t_{i},s_{i})$), with
mean $\mu(\mathbf{t},\mathbf{c})$ and variance $\nu(\mathbf{t},\mathbf{c})$
given by Eqs.~(\ref{eq:mean_t}) and (\ref{eq:variance_t}), respectively.
And if $E(\mathbf{t},\mathbf{s},\mathbf{c})$ is Gaussian distributed,
then the limiting distribution of $\min_{k=1}^{M}E(\mathbf{t},\mathbf{s}^{(k)},\mathbf{c})$
as $M\to\infty$ will be the Gumbel distribution, which has cumulative
distribution function
\begin{equation}
P_{G}(\min_{k}E^{(k)}<E)=1-\exp\left[-\exp\left(\frac{E-a_{M}(\mathbf{t},\mathbf{c})}{b_{M}(\mathbf{t},\mathbf{c})}\right)\right],\label{eq:CDF_Gumbel}
\end{equation}
where\begin{equation}
a_{M}(\mathbf{t},\mathbf{c})=\mu(\mathbf{t},\mathbf{c})-\sqrt{2\nu(\mathbf{t},\mathbf{c})}\alpha,\label{eq:a_M-1}
\end{equation}
with $\alpha=\sqrt{\log M}-\frac{1}{4\sqrt{\log M}}\big(\log\log M+\log4\pi\big)+\mathcal{O}((\log M)^{-3/2})$
(depending only on $M$), and\begin{equation}
b_{M}(\mathbf{t},\mathbf{c})=\sqrt{\frac{\nu(\mathbf{t},\mathbf{c})}{2\log M}}+\mathcal{O}((\log M)^{-3/2}\big).\label{eq:b_M}
\end{equation}
The peak of this distribution is $\rho_{G}(\mathbf{t},\mathbf{c})=a_{M}(\mathbf{t},\mathbf{c})$,
and its variance is $\nu_{G}(\mathbf{t},\mathbf{c})=\frac{\pi^{2}}{6}b_{M}^{2}(\mathbf{t},\mathbf{c})$.
In the main text of this paper, we used the leading-order value of
$\alpha$, $\alpha=\sqrt{\log M}$ (see after Eq.~(\ref{eq:a_M})).

Using Bayes' rule, the posterior distribution of TCR sequences surviving
selection, $P(\mathbf{t}\big|\text{post},\mathbf{c})$, is given by
\begin{widetext}
\begin{align}
P(\mathbf{t}\big|\text{post},\mathbf{c}) & =\frac{P(\text{post}\big|\mathbf{t},\mathbf{c})P_{\text{pre}}(\mathbf{t})}{\sum_{\mathbf{t}}P(\text{post}\big|\mathbf{t},\mathbf{c})P_{\text{pre}}(\mathbf{t})}\nonumber \\
 & \propto\big[P_{G}(E<E_{p})-P_{G}(E<E_{n})\big]\times P_{\text{pre}}(\mathbf{t})\nonumber \\
 & =\left\{\exp\left[-\exp\left(\frac{E_{n}-a_{M}(\mathbf{t},\mathbf{c})}{b_{M}(\mathbf{t},\mathbf{c})}\right)\right]-
 \exp\left[-\exp\left(\frac{E_{p}-a_{M}(\mathbf{t},\mathbf{c})}{b_{M}(\mathbf{t},\mathbf{c})}\right)\right]\right\}\times P_{\text{pre}}(\mathbf{t}).\label{eq:p(t|sel)}
\end{align}
\end{widetext}
While this is exact within the Gaussian approximation,
it is not immediately obvious how to make progress quantifying the
enrichment of post-selection TCR amino acids by direct marginalization,
i.e. $P(t_{i}\big|\text{post},\mathbf{c})=\sum_{\{t_{j}\}_{j=1\dots N\backslash i}}P(\mathbf{t}=(t_{1}\dots t_{N})\big|\text{post},\mathbf{c})$.

Reference~\cite{2009-Kosmrlj} made progress in the limit of $N,M\to\infty$
(keeping $N\propto\log M$), when the extreme value distribution concentrates
around its peak, $\rho_{G}(\mathbf{t},\mathbf{c})$, which lies somewhere
between $E_{n}$ and $E_{p}$. Now, we ask: what is the probability
distribution that minimizes the relative entropy (or Kullback\textendash Liebler
divergence) to the pre-selection distribution $P_{\text{pre}}(\mathbf{t})$,
given the constraint that $\rho_{G}(\mathbf{t},\mathbf{c})$ lies
between $E_{n}$ and $E_{p}$? The answer is
\begin{equation}
P(\mathbf{t}\big|\text{post},\mathbf{c})=\frac{1}{Z_{\beta,\mathbf{c}}}e^{-\beta(\mu(\mathbf{t},\mathbf{c})-\sqrt{2\nu(\mathbf{t},\mathbf{c})}\alpha)}P_{\text{pre}}(\mathbf{t}),\label{eq:P(t|sel)_2}
\end{equation}
where $Z_{\beta,\mathbf{c}}$ ensures normalization, and we have used
Eq.~(\ref{eq:a_M-1}). Here, $\beta$ is a Lagrange multiplier constraining
the value of $\langle\mu(\mathbf{t},\mathbf{c})-\sqrt{2\nu(\mathbf{t},\mathbf{c})}\alpha\rangle\equiv\sum_{\mathbf{t}}\big(\mu(\mathbf{t},\mathbf{c})-\sqrt{2\nu(\mathbf{t},\mathbf{c})}\alpha\big)P(\mathbf{t}\big|\text{post},\mathbf{c})$
to lie between $E_{n}$ and $E_{p}$. The optimal value of $\beta$,
$\beta^{*}$, is the one as close to 0 as possible that satisfies
this constraint. The mapping from hard constraints on the extreme value
to a constraint on its mean is analogous to that from the microcanonical
to the canonical ensemble in the thermodynamic limit~\cite{2009-Kosmrlj}.

The marginal distribution of amino acid $t_{i}$ at position $i$
of post-selection TCR sequences, $P(t_{i}\big|\text{post},\mathbf{c})$,
may be obtained from Eq.~(\ref{eq:P(t|sel)_2}) by taking a sum over
$20^{N-1}$ terms. However, in the $N\to\infty$ limit, $\nu(\mathbf{t},\mathbf{c})$
self-averages, i.e. $\sum_{i=1}^{N}\nu(t_{i},c_{i})\to\sum_{i=1}^{N}\langle\nu(t,c_{i})\rangle$;
thus, performing the double Legendre transform on $\rho_{G}(\mathbf{t},\mathbf{c})$
w.r.t. $\nu(\mathbf{t},\mathbf{c})$ (equivalently Hamiltonian minimization
\cite{1983-Kardar}), and replacing $\nu(\mathbf{t},\mathbf{c})$
by its self-averaged value $\sum_{i=1}^{N}\langle\nu(t,c_{i})\rangle$,
Eq.~(\ref{eq:P(t|sel)_2}) factorizes into Eq.~(\ref{eq:factor_CaseAB})
of the main text, where $\gamma$ (given by Eq.~(\ref{eq:gamma_CaseAB}))
is the conjugate variable to $\nu(\mathbf{t},\mathbf{c})$. After
the Legendre transforms w.r.t. $\nu(\mathbf{t},\mathbf{c})$, $\langle\rho_{G}(\mathbf{t},\mathbf{c})\rangle$
becomes Eq.~(\ref{eq:mu_G_CaseAB}). How $\beta^{*}$ is found in
practice is described after Eq.~(\ref{eq:mu_G_CaseAB}) of the main
text.
\end{document}